\pdfoutput=1
\documentclass[11pt,twoside,a4paper,cmspaper,final,collab]{cms-tdr}
\def\svnVersion{4c981ff-D}\def\svnDate{2021/07/23}\def\cmsCernNoTag{CERN-EP-2021-049}\def\cmsCernDate{\today}\def\cmsMessage{"Published in Physics Letters B as \href{http://dx.doi.org/10.1016/j.physletb.2021.136535}{\doi{10.1016/j.physletb.2021.136535}.}"}
\begin{document}\cmsNoteHeader{B2G-20-005}

\hyphenation{had-ron-i-za-tion}
\hyphenation{cal-or-i-me-ter}

\renewcommand{\ttbar}{\ensuremath{\PQt\PAQt}\xspace}
\newcommand{\msd}{\ensuremath{m_{\mathrm{SD}}}\xspace}
\newcommand{\Wprime}{\PWpr\xspace}
\newcommand{\mtb}{\ensuremath{m_{\PQt\PQb}}\xspace}
\newcommand{\pb}{\ensuremath{\,\text{pb}}\xspace}
\newcommand{\DeepJet}{\textsc{DeepJet}\xspace}
\newcommand{\DeepAKX}{\textsc{DeepAK8}\xspace}
\newcommand{\Rpf}{\ensuremath{R_{\text{p/f}}}\xspace}
\newcommand{\sigWp}{\ensuremath{{\sigma}_{\Wprime}}\xspace}

\cmsNoteHeader{B2G-20-005} 
\title{Search for \texorpdfstring{$\Wprime$}{\Wprime} bosons decaying to a top and a bottom quark at \texorpdfstring{$\sqrt{s} = 13\TeV$}{sqrt(s) = 13 TeV} in the hadronic final state}

\date{\today}

\abstract{
   A search is performed for $\Wprime$ bosons decaying to a top and a bottom quark in the all-hadronic final state, in proton-proton collisions at a center-of-mass energy of 13\TeV. 
   The analyzed data were collected by the CMS experiment between 2016 and 2018 and correspond to an integrated luminosity of 137\fbinv.
   Deep neural network algorithms are used to identify the jet initiated by the bottom quark and the jet containing the decay products of the top quark when the $\PW$ boson from the top quark decays hadronically.
   No excess above the estimated standard model background is observed. 
   Upper limits on the production cross sections of $\Wprime$ bosons decaying to a top and a bottom quark are set.
   Both left- and right-handed $\Wprime$ bosons with masses below 3.4\TeV are excluded at 95\% confidence level, 
   and the most stringent limits to date on $\Wprime$ bosons decaying to a top and a bottom quark in the all-hadronic final state are obtained.
}

\hypersetup{
pdfauthor={CMS Collaboration},%
pdftitle={Search for W-prime bosons decaying to a top and a bottom quark at sqrt(s) = 13 TeV in the hadronic final state},%
pdfsubject={CMS},
pdfkeywords={CMS, W' search, top resonances, boosted top}}

\maketitle

\section{Introduction}\label{sec:intro}

The CERN LHC has strengthened the validity of the standard model (SM) of particle physics by providing a large volume of data to be compared with theoretical predictions. 
The existence of new physics beyond the SM, however, is needed in order to explain several observed phenomena, including the indications for the existence of dark matter, the origin of nonzero neutrino masses, and the baryon asymmetry of the universe.
Also, an explanation of the fine tuning required for the insensitivity of the Higgs boson mass to quantum corrections in the SM is one of the important theoretical quests in particle physics.
Extensions of the SM, conceived to overcome these limitations, include theories proposing
a new spin-1 gauge boson $\Wprime$,  a color singlet with an electric charge of $\pm 1$.
The $\Wprime$ boson appears, for example, in left-right symmetric models~\cite{PhysRevD.11.2558,PhysRevD.36.878}, in models with extra spatial dimensions~\cite{Burdman:2006gy}, and in little Higgs models~\cite{PhysRevD.10.275}.
Several of these models predict $\Wprime$ bosons having either right-handed or left-handed charged current interactions. 
In the latter case, interference with SM single top quark production can be present. 
A model-independent description of such processes can be found in~\cite{Boos:2006xe}.
Some theoretical models, for example~\cite{Hill:1994hp,Muller:1996dj,Abdullah:2018ets}, assume a preferential coupling of a $\Wprime$ boson to the third-generation fermions, which motivates the search for a $\Wprime$ boson decaying to a top and a bottom quark.

The first searches for a $\Wprime$ boson decaying to a top and a bottom quark were conducted
by the CDF and D0 experiments 
at the Tevatron~\cite{Aaltonen:2015xea,Abazov:2011xs} in proton-antiproton collisions, followed by those of the ATLAS and CMS experiments at the LHC using data from proton-proton ($\Pp\Pp$) collisions~\cite{Chatrchyan:2012gqa,Chatrchyan:2014koa,Aad:2014xea,Khachatryan:2015edz,Aaboud:2018juj,Sirunyan:2017vkm}.
In the previous searches for $\Wprime$ bosons in all-hadronic decay modes, 
the CMS experiment excluded right-handed $\Wprime$ bosons with masses less than 2\TeV at 95\% confidence level (\CL)  at $\sqrt{s}=8\TeV$, using data corresponding to an integrated luminosity of 19.7\fbinv~\cite{Khachatryan:2015edz}.
The ATLAS experiment excluded right- (left-)handed $\Wprime$ bosons below 3.0 (2.9)\TeV at $\sqrt{s}=13\TeV$ 
using data collected in 2015 and 2016, corresponding to an integrated luminosity of  36.1\fbinv~\cite{Aaboud:2018juj}.
The search performed by the CMS experiment on a similar data set, considering only
leptonic final states from the $\PQt\to\PQb\PW\to\PQb\Pl\PGn$ decay chain, excluded right- (left-)handed $\Wprime$ bosons of masses roughly below 3.6 (3.4)\TeV  at 95\% \CL~\cite{Sirunyan:2017vkm}.

In this letter, we search for a $\Wprime$ boson decaying to a top and a bottom quark in the all-hadronic final state, where the signature is an excess of events over a smoothly falling background in the invariant mass spectrum of top and bottom quark candidates (\mtb) in the range 1--4\TeV.

The main SM background processes from LHC $\Pp\Pp$ collisions that can mimic the final state sought in this search are the production of multijet events due to quantum chromodynamics (QCD) interactions, the production of a top quark-antiquark pair ({\ttbar}), and the electroweak production of a single top quark associated with a bottom quark or a \PW boson.
As none of the background processes involve a heavy resonance in the $s$ channel, they result in smoothly falling \mtb spectra.

Most of the top quarks in decays of $\Wprime$ bosons with masses greater than 1\TeV have large transverse momentum (\pt), and their subsequent decay products are clustered into a single jet of particles. 
This requires the use of techniques based on jet substructure~\cite{Larkoski:2017jix,Asquith:2018igt} and machine learning~\cite{Kasieczka:2019dbj} algorithms for the identification of the jets due to top quark decay (\PQt tagging)
that significantly reduce the background from multijet production in QCD in the case of an all-hadronic final state.
The jets identified as coming from such highly Lorentz-boosted top quarks are referred to as {\PQt}-tagged jets. 
This search makes use of the latest advancements in CMS in bottom-quark tagging~\cite{Bols:2020bkb} as well as top-quark tagging~\cite{Sirunyan:2020lcu} involving a deep neural network (DNN).
The study is based on data corresponding to an integrated luminosity of 137\fbinv collected by the CMS experiment in $\Pp\Pp$  collisions at $\sqrt{s}=13\TeV$ from 2016 to 2018.

Tabulated results are provided in HEPDATA~\cite{hepdata}.

\section{The CMS detector} \label{sec_CMS}

The central feature of the CMS apparatus is a superconducting solenoid of 6\unit{m} internal diameter, providing a magnetic field of 3.8\unit{T}. Within the solenoid volume are a silicon pixel and silicon strip tracker, a lead tungstate crystal electromagnetic calorimeter (ECAL), and a brass-and-scintillator hadron calorimeter (HCAL), each composed of a barrel and two endcap sections. Forward calorimeters extend the coverage in pseudorapidity ($\eta$) provided by the barrel and endcap detectors. Muons are detected in gas-ionization chambers embedded in the steel flux-return yoke outside the solenoid. A more detailed description of the CMS detector, together with a definition of the coordinate system used and the relevant kinematic variables, can be found in Ref.~\cite{CMS_ex}.

The silicon tracker measures charged particles within the range $\abs{\eta} < 2.5$. 
It consists of 1440 silicon pixel and 15\,148 silicon strip detector modules. 
In 2017, an additional layer was added in both the barrel and endcap regions of the pixel detector and
the number of silicon pixel modules increased to 1856.
For nonisolated particles with $1 < \pt < 10\GeV$ and $\abs{\eta} < 1.4$, the track resolutions are typically 1.5\% in \pt and 25--90\mum and 25--75\mum in the transverse impact parameter in 2016 and 2017 onwards, respectively, whereas the resolution in the longitudinal impact parameter is 45--150\mum~\cite{TRK-11-001,CMS-DP-2020-049}.

Events of interest are selected using a two-tiered trigger system~\cite{Khachatryan:2016bia}. The first level (L1), composed of custom hardware processors, uses information from the calorimeters and muon detectors to select events at a rate of around 100\unit{kHz}. The second level, known as the high-level trigger (HLT), consists of a farm of processors running a version of the full event reconstruction software optimized for fast processing that reduces the event rate to around 1\unit{kHz} before data storage.

\section{Object reconstruction}\label{sec:objects}

The CMS particle-flow algorithm~\cite{PFA} aims to reconstruct and identify individual particles in an event with an optimized combination of information from the various elements of the CMS detector. 
The reconstructed vertex with the largest value of summed physics-object $\pt^2$ is taken to be the primary $\Pp\Pp$ interaction vertex. 
The energy of photons is obtained from ECAL clusters that have no associated track.
The energy of electrons is determined from a combination of the electron momentum at the primary interaction vertex as determined by the tracker, the energy of the corresponding ECAL cluster, and the energy sum of all bremsstrahlung photons spatially compatible with originating from the electron track. 
The energy of muons is obtained from the curvature of the corresponding track as determined using the tracker and the muon system. The energy of charged hadrons is determined from a combination of their momentum measured in the tracker and the matching ECAL and HCAL energy deposits, corrected for the response function of the calorimeters to hadronic showers. Finally, the energy of neutral hadrons is obtained from the corresponding corrected ECAL and HCAL energy deposits.

For each event, hadronic jets are clustered from these reconstructed particles
(particle-flow candidates) using the infrared- and collinear-safe anti-\kt algorithm~\cite{anti-kT} with distance parameters $0.4$ (AK4 jets) and $0.8$ (AK8 jets), as implemented in the \FASTJET package~\cite{Cacciari:2011ma}. 
The AK4 and AK8 jets are used to identify the bottom quark and the hadronically-decaying top quark, respectively, from a $\Wprime$ boson decay.
The distance between two particles in the $\eta$-$\phi$ plane, where $\phi$ is azimuthal angle in radians, is defined as $\Delta R = \sqrt{\smash[b]{(\Delta\eta)^2+(\Delta\phi)^2}}$.
The jet momentum is determined as the vectorial sum of all particle momenta in the jet, and is found from simulation to be, on average, within 5--10\% of the momentum of the particle-level jets reconstructed using stable particles (lifetime $>30\unit{ps}$), excluding neutrinos, for $\pt>50\GeV$ and rapidity $\abs{y}<2.5$.
Additional $\Pp\Pp$ interactions within the same or nearby bunch crossings (pileup) can result in additional tracks and calorimetric energy depositions, increasing the apparent jet momentum. 
To mitigate this effect, 
tracks identified as originating
from pileup vertices are discarded before jet reconstruction. 
For AK4 jets, an offset correction~\cite{Cacciari:2007fd} is applied to correct for remaining pileup contributions~\cite{CMS_JES_8TeV}. 
For AK8 jets, the pileup per particle identification (PUPPI) algorithm~\cite{Bertolini:2014bba} is used to mitigate the effect of pileup at the reconstructed particle level.  
It has been shown that the PUPPI algorithm improves the resilience of jet substructure observables against pileup~\cite{Sirunyan:2020foa}.

Additional selection criteria are applied to each jet to remove those potentially dominated by instrumental effects or reconstruction failures~\cite{CMS-PAS-JME-16-003}. 
These criteria are the following:
the fraction of the jet energy
carried by neutral hadrons and photons should be less than 90\%, the jet should have at least two constituents, and at least one of those should be a charged hadron.  
These requirements remove approximately 0.5\% of jets selected for analysis, with negligible loss of genuine jets.

Jet energy corrections are derived from simulation so that the average measured energy of jets is the same as that of the corresponding particle-level jets. 
Measurements of the momentum balance in dijet, photon+jet, $\PZ+$jet, and multijet events are used to determine any residual differences between the jet energy scale (JES) in data and simulation, and appropriate corrections are made~\cite{CMS_JES_8TeV}.
Jet energy correction factors are derived using this methodology for both
AK4 and AK8 jets.
The jet energy resolution (JER) is obtained from a dijet balance technique~\cite{CMS-DP-2018-028}.
To match the JER in data and simulation, an energy smearing is added to the AK4 and AK8 jets in simulation.

\subsection{Identification of jets from bottom quarks}

A DNN-based tagger, {\DeepJet}~\cite{Bols:2020bkb}, is used for the \PQb tagging of AK4 jets, utilizing information from the tracks, neutral particles, and the secondary vertices within the jet.
This tagger also provides multiple outputs, such as whether the input jet is consistent with a jet initiated from one or more \PQb quarks, one or more \PQc quarks, light quarks, or gluons.

The thresholds used for the \DeepJet \PQb tagger correspond to a mistag rate for jets initiated by light quarks or gluons at a $\pt>500\GeV$ of approximately 5\% for data and simulated samples in 2016 and 1\% in 2017 and 2018.
This choice of threshold corresponds to an efficiency of approximately 75 (60)\%  at $\pt=500\GeV$ and 65 (50)\% at $\pt=1000\GeV$ for jets initiated by \PQb quarks in the barrel (endcap) region. 
To match the shape of the \DeepJet discriminator in data and simulation, 
corrections as a function of the \pt and $\eta$ of AK4 jets, 
derived using samples enriched in dileptonic \ttbar events for the \PQb and \PQc quark-initiated jets, 
and $\PZ+$jets events for the jets initiated by light quarks and gluons, 
are applied in simulation. 
The \PQb tagging performance is better
in 2017--2018 than in 2016
because of the addition in 2017
of new layers in the pixel detector of the CMS tracker close to the interaction point, in both the barrel and endcap regions.

\subsection{Identification of jets from top quarks} 

The top quark arising from the decay of a heavy $\Wprime$ boson has a large Lorentz boost, and its decay products are expected to be captured within a jet with a large distance parameter.
Hence AK8 jets are used to identify {\PQt}-tagged jets. 
The key observables for the selection of AK8 jets resulting from top quark decay are as follows:
\begin{itemize}

\item Groomed jet mass: grooming is a procedure for the removal of soft radiation clustered into the jet, which mitigates effects from initial- and final-state radiation, underlying event~\cite{PhysRevD65092002}, and pileup interactions. Grooming results in a proportionally larger reduction in the mass of jets from light quarks or gluons with respect to genuine top quarks.

\item Substructure of the jet: 
three dense clusters of energy are expected inside the jet, associated with the \PQb quark and the hadronic decay of the \PW boson, whereas a jet originated from a light quark or gluon is generally characterized by a single cluster of energy.

\item Displaced vertices: the presence of at least one displaced vertex is expected from the decays of \PQb hadrons.

\end{itemize}

The soft drop algorithm,  a generalization of the modified mass drop algorithm~\cite{Butterworth:2008iy,Dasgupta:2013ihk},
with angular exponent $\beta = 0$, soft cutoff threshold $z_{\text{cut}} = 0.1$, and characteristic radius $R_{0} = 0.8$~\cite{Larkoski:2014wba} 
is used to groom the AK8 jets,
and the corresponding groomed mass, known as the soft-drop mass~(\msd), is required to be within a window of 105--210\GeV for a jet to be {\PQt}-tagged.
In this algorithm, the constituents of the AK8 jets are reclustered using the Cambridge--Aachen algorithm~\cite{Dokshitzer:1997in,Wobisch:1998wt} and  
the relative \pt between the successive clusters of particles merged during the jet clustering is checked to remove soft, wide-angle particles from the jet.

The latter two of the features mentioned above are incorporated in a DNN designed to identify a jet arising from top quark decay.
The DNN based \PQt tagging algorithm exploited in this study, henceforth referred to as the \DeepAKX tagger, has been studied extensively in CMS~\cite{Sirunyan:2020lcu}. 
This algorithm uses up to a hundred particles (selected in descending order of \pt) in an AK8 jet, incorporating six kinematic variables (\pt, $\eta$, $\phi$, $\Delta R$ from the jet axis, $\Delta R$ from the axes of two soft-drop subjets) for each of the particles and also exploits the features of the tracks (quality, displacement, etc.), and properties of secondary vertices associated with the jet.
A relative score is assigned to the jet by the \DeepAKX tagger 
specifying how likely 
the jet is to have originated from the decay products of a top quark rather than from a light quark or gluon.
A recursive neural network based approach is used to decorrelate the tagger performance from the jet mass.

The threshold used on the \DeepAKX tagger score corresponds to a rate of incorrectly tagging jets originated from light quarks or gluons, called mistag rate of 0.5\%. 
This choice of threshold corresponds to an efficiency of approximately 35--45\%, in the phase space of this analysis, to identify the jets initiated by top quarks.
The efficiency of the \DeepAKX tagger is measured in single-muon events enriched with semileptonic \ttbar production, and increases with jet \pt.
Corrections based on the \pt of AK8 jets are applied in simulation to match the efficiency of the \PQt tagging algorithm in data~\cite{Sirunyan:2020lcu}.

\section{Data and simulated samples}
\label{sec:samples}

The data used in this search are from $\Pp\Pp$ collisions at $\sqrt{s}=13\TeV$ collected by the CMS experiment from 2016 to 2018, corresponding to an integrated luminosity of 137\fbinv.

Monte Carlo simulations are used to model the $\Wprime$ boson signal and the background sources relevant to this analysis.
Background estimates for \ttbar events are taken primarily from simulation, but also include a correction derived from data.
Background estimates for multijet production are taken entirely from data; simulated multijet samples are used for cross-checks.
The smallest background considered is from single top quark events, and it is estimated purely from the simulation.

The signal samples are generated at leading order (LO) using the \COMPHEP~{v4.5.2} generator~\cite{Boos:2004kh}. 
Signal samples are generated separately for left- and right-handed $\Wprime$ bosons 
with masses between 1--4\TeV in steps of 100\GeV.
The width of the $\Wprime$ boson in all of the generated samples is $\sim$3\%~\cite{Boos:2006xe}.
The cross sections of the signal samples are scaled to next-to-leading-order (NLO) accuracy using a $K$ factor of 1.25~\cite{Sullivan:2002jt,Duffty:2012rf}.
The value of the ${\Lambda}_{\text{QCD}}$ parameter is taken to be $165.2\MeV$ in the signal samples.

The \POWHEG~2.0 generator~\cite{Nason:2004rx,Powheg_ref,PowhegBox_ref} is used to generate \ttbar events at NLO in perturbative QCD~\cite{Frixione:2007nw}. 
For the normalization of the \ttbar sample, the production cross section calculated at next-to-next-to-leading order (NNLO) with the resummation of soft gluons at next-to-next-to-leading logarithmic precision~\cite{Czakon:2011xx} is used. 
Event generation for the production of a single top quark in the $t$ channel~\cite{fourfiveflavorschemes} and in association with a \PW boson~\cite{Re:2010bp}, is performed with the \POWHEG~2.0 generator as well. 
The sample of events with a single top quark produced in association with a \PW boson~\cite{Re:2010bp} is normalized to the NNLO cross section~\cite{Kidonakis:2010ux}.
Events with the production of a single top quark in the $s$ channel are generated at NLO using \MGvATNLO~\cite{Alwall:2014hca}, where version 2.2.2 is used for 2016 and version 2.4.2 is used for the 2017 and 2018 data-taking eras, and top quark decays are simulated with {\sc madspin}~\cite{Artoisenet:2012st}.
The QCD multijet events are produced with the \MGvATNLO generator at LO with up to four outgoing partons in the final state.

Simulated \ttbar and single top quark samples in 2016 make use of NNPDF3.0~\cite{Ball:2014uwa} NNLO parton distribution functions (PDFs), with the strong coupling constant $\alpS (M_{\mathrm{\PZ}})$ set to $0.118$, to describe the momentum distribution of partons inside the colliding protons.
The NNPDF3.0 LO PDFs and  are used in producing simulated multijet samples in 2016.
The NNPDF3.1~\cite{Ball:2017nwa} NNLO PDFs are used to simulate \ttbar, single top quark, and multijet samples in 2017 and 2018.
Simulated signal samples use the CTEQ6L1~\cite{Pumplin:2002vw} LO PDF set.

Generated partons undergo parton showering and hadronization using \PYTHIA~{v8.212}~\cite{Sjostrand:2014zea}.
For the simulated multijet sample,
the matching of \PYTHIA to \MGvATNLO is performed in the MLM~\cite{Alwall:2014hca} schemes.
In the case of the samples with \ttbar and single top quark production, \POWHEG and \MGvATNLO are matched to \PYTHIA using the FxFx~\cite{Frederix:2012ps} scheme.
The underlying event activity in each sample, except for the QCD multijet and $\Wprime$ boson signals in the 2016 era, is simulated using the CP5 tune, which is derived by tuning the model parameters for multiple parton interactions in \PYTHIA using minimum bias data collected by the CMS experiment~\cite{Sirunyan:2019dfx}. 
For the simulated QCD multijet events and $\Wprime$ boson signals in the 2016 era, the underlying event tune is CUETP8M1~\cite{CUETP8M1_CMS}.
For all samples, in order to match the pileup conditions in data and simulation, a weighting is performed in simulation based on the value of the total inelastic cross section, which is taken to be 69.2\unit{mb}~\cite{Sirunyan:2018nqx}. 
The generated samples are processed through the CMS detector simulation based on {\GEANTfour}~\cite{Agostinelli:2002hh}, using the same reconstruction algorithms as data.

\section{Event selection}
\label{sec:event_sel}

The trigger criteria chosen in this analysis exploit the large amount of hadronic activity expected in signal events.
At L1, a combination of several criteria consisting of requirements on the \pt of AK4 jets or \HT, defined as the scalar \pt sum of all of the AK4 jets in the event with $\pt>30\GeV$ and $\abs{\eta}<3.0$, is applied to select the events.
In the HLT, a set of trigger conditions is applied, selecting events that meet at least one of the following requirements:
\begin {itemize}
\item there is at least one AK4 or AK8 jet above a \pt threshold of 450 (500)\GeV for the data-taking period 2016 (2017 and 2018);
\item there is at least one AK8 jet that has a \pt greater than 360 or 420\GeV in the data-taking periods of 2016 or 2017 and onwards, respectively, and a groomed mass of at least 30\GeV, where a trimming~\cite{Krohn:2009th} algorithm is used for jet grooming;
\item \HT is greater than a threshold that varied between 800 and 1050\GeV depending on the data-taking period and instantaneous luminosity;
\item the scalar \pt sum of all of the AK8 jets with $\pt > 150 \ (200)\GeV$ and $\abs{\eta}<2.5$ is greater than 700 (900)\GeV in 2016 (2017 and 2018), and at least one of the AK8 jets has a groomed mass of at least 50\GeV.
\end{itemize}
Events with at least one isolated electron or muon with $\pt>30\GeV$ are rejected, 
where the condition for electron and muon identification corresponds to approximately 90 and 95\% efficiency for a genuine electron or muon, respectively.
Events with at least one AK8 and one AK4 jet, both with $\pt>550\GeV$ and $\abs{\eta}<2.4$,
where the AK8 and AK4 jets are separated by $\Delta R \geq 1.2$, are considered for the analysis.  
The AK8 jet with the highest \PQt tagging score is taken as the top quark candidate jet.
As the top quark and bottom quark from $\Wprime$ boson decays are expected to be produced in a back-to-back topology,
the AK4 jet with the highest \pt which satisfies $\Delta\phi>\pi/2$ with respect to the top quark candidate jet is taken to be the bottom quark candidate jet.
If an AK8 jet is present within $\Delta R < 0.4$ of the \PQb quark candidate jet, it is referred to as the AK8 jet associated with the \PQb quark candidate jet.
The \pt thresholds of the top quark and bottom quark candidate jets are chosen such that the triggers used are more than 99\% efficient for the selected events. 
The efficiency of the triggers has been measured in data and simulation with respect to a reference trigger, which requires the event to have at least one muon, and differences between the two are found to be within 0.1\% in the phase space of selected events. 
Therefore, no correction is applied to the simulated samples.

A further selection criterion is applied to reduce the contamination from the \ttbar background. 
After the top and bottom quark candidate jets are selected, 
if the AK8 jet associated with the bottom quark candidate exists and it has \msd greater than 60\GeV, the event is discarded. 
This requirement is imposed to reject bottom quark jets from the hadronic decay chain of top quarks.

\section{Event categorization and background estimation}
\label{sec:bkg}

After applying the event selection described in Section~\ref{sec:event_sel}, 
events are further divided into regions, 
depending on whether the top or bottom quark candidate jets pass or fail the tagging requirements, 
for the estimation of multijet background. 
The following naming convention is used for the phase space division:
\begin{itemize}
\item {\PQt}: top quark candidate AK8 jet;
\item {\PQb}: bottom quark candidate AK4 jet;
\item \PQt tagging pass or fail: \PQt passes or fails the threshold on the \PQt tagging score;
\item \PQb tagging pass or fail: \PQb passes or fails the threshold on the \PQb tagging score.
\end{itemize}
The signal region SR is defined by requiring the top quark candidate jet to pass both the \msd and the \PQt tagging score requirement and the bottom quark candidate jet to pass the threshold on the \PQb tagging score.
The control region SR$^\prime$ is defined with the same conditions on the top quark candidate jet as in the SR, 
but the bottom quark candidate jet is required to fail the requirement on the \PQb tagging score. 
The control regions CR1 and CR1$^\prime$ are similar to the regions SR and SR$^\prime$, respectively, apart from the \msd requirement on the top quark candidate jet, which is changed as indicated in Table~\ref{tab:Region}.
The validation region VR and other control regions VR$^\prime$, CR2 and CR2$^\prime$ are defined by the same criteria used to build SR, SR$^\prime$, CR1, and CR1$^\prime$, respectively, and differ only in the \PQt tagging condition. 
An overview of the regions used in the analysis is given in Table~\ref{tab:Region}. 
\begin{table*}[htbp]
\centering
\topcaption{
Regions of parameter space used in the analysis. 
The \msd and \PQt tagging refer to the soft drop mass and the {\DeepAKX} {\PQt}-tagger score requirements of the top quark candidate AK8 jet. The \PQb tagging refers to the {\DeepJet} {\PQb}-tagger score requirement of the bottom quark candidate AK4 jet.
}
\renewcommand{\arraystretch}{1.2}
\begin{tabular}{ m{1.cm}  m{2.75cm} m{1.8cm}  m{1.8cm} m {5.75cm}} 
Region & \hspace{0.5cm}\msd &  \PQt tagging & \PQb tagging &  Purpose \\
\hline
SR     &  $\in [105,210]\GeV$  & pass  & pass  & Signal extraction\\
SR$^\prime$     &  $\in [105,210]\GeV$  & pass  & fail  & Multijet bkg. estimation in SR\\
VR     &  $\in [105,210]\GeV$  & fail  & pass  & Validation of bkg. estimation\\
VR$^\prime$     &  $\in [105,210]\GeV$  & fail  & fail  & Multijet bkg. estimation in VR\\
CR1    &  $<$105\GeV		   & pass  & pass & Extrapolation of multijet bkg. \\
CR1$^\prime$    &   $<$105\GeV          & pass  & fail & \hspace{0cm}from SR$^\prime$ to SR\\
CR2    &  $<$105\GeV		   & fail  & pass & Extrapolation of multijet bkg.\\
CR2$^\prime$    &   $<$105\GeV          & fail  & fail  & \hspace{0cm}from VR$^\prime$ to VR\\
\end{tabular}
\label{tab:Region}
\end{table*}

The criteria on \msd and \PQt tagging score of the top quark candidate jet and \PQb tagging score of the bottom quark candidate jet
are chosen to achieve maximum sensitivity to a $\Wprime$ boson signal in the SR, where the multijet background constitutes 85--90\% of the total background, whereas \ttbar and single top quark production contribute 5--8 and 2--5\%, respectively.
For a right-handed $\Wprime$ boson of mass 2\TeV, the signal selection efficiency, 
defined as the fraction of the simulated events with the production of a $\Wprime$ boson decaying into a top and a bottom quark in the all-hadronic final state falling within the SR,
is approximately 8\% in 2016 and 9\% in 2017 and 2018.
The efficiency for selecting the signal events with a right-handed $\Wprime$ boson of mass 4\TeV is about 5\% in all years.
For left-handed $\Wprime$ bosons, the signal selection efficiency is approximately 5\% for a 2\TeV resonance mass and decreases to 0.1\% for a 4\TeV resonance mass.
The large difference between the signal selection efficiency for left-handed and right-handed $\Wprime$ bosons, especially for high resonance masses, is due to the interference with single top quark production in the case of left-handed $\Wprime$ bosons, which results in a larger number of events at low energy that tend to fall outside the acceptance of the SR.

The control regions CR1 and CR1$^\prime$ are used to derive the \PQb tagging pass-to-fail ratio (\Rpf) of the \PQb quark candidate jet.
The ratio \Rpf obtained from  CR1 and CR1$^\prime$, and the event yield in the control region SR$^\prime$ are used to estimate the multijet background in the SR. 
The technique used to estimate the multijet background is cross-checked in the VR, where the multijet background is computed using the regions CR2, CR2$^\prime$, and VR$^\prime$.
The parton flavor composition of the \PQb quark candidate jet has been studied using simulated samples and has been found to be comparable between the SR and the CR1 used to derive \Rpf. The same comparability has been verified for VR and CR2.

The ratios \Rpf are obtained by dividing the \mtb spectrum obtained in CR1 by that from CR1$^\prime$, and similarly the  \mtb spectrum in CR2 by that in CR2$^\prime$, as shown in Eq.~\eqref{Eq:pfrat}:
\begin{linenomath}
\begin{equation}
\begin{split}
R^{1}_{\text{p/f}} (\mtb) = \frac{\mathrm{CR}1}{\mathrm{CR}1^\prime}, \\
R^{2}_{\text{p/f}} (\mtb) = \frac{\mathrm{CR}2}{\mathrm{CR}2^\prime}.
\end{split}
\label{Eq:pfrat}
\end{equation}
\end{linenomath}
The ratios $R^{1}_{\text{p/f}}$ and $R^{2}_{\text{p/f}}$, obtained as functions of \mtb, are parameterized using a second-order polynomial. 
The ratios \Rpf are also fitted with a bifurcating function, defined in Eq.~\eqref{Eq:BiFunc}, to estimate the systematic uncertainty associated with the choice of the parameterization.
\begin{linenomath}
\ifthenelse{\boolean{cms@external}}
{
\begin{multline}
\begin{aligned}
f_{\text{p/f}}(\mtb) =
 \begin{cases}
 	a_{1} + a_{2} \left(\mtb-a_{0}\right) \\ + a_{3} {\left(\mtb-a_{0}\right)}^{2}, & \mtb<a_{0}\\
 	a_{1} + a_{2} \left(\mtb-a_{0}\right) \\ + a_{4} {\left(\mtb-a_{0}\right)}^{2}, & \mtb{\geq}a_{0}
 \end{cases}.
\end{aligned}
\label{Eq:BiFunc}
\end{multline}
}
{
\begin{equation}
\begin{aligned}
f_{\text{p/f}}(\mtb) =
 \begin{cases}
 	a_{1} + a_{2} \left(\mtb-a_{0}\right) + a_{3} {\left(\mtb-a_{0}\right)}^{2}, & \mtb<a_{0}\\
 	a_{1} + a_{2} \left(\mtb-a_{0}\right) + a_{4} {\left(\mtb-a_{0}\right)}^{2}, & \mtb{\geq}a_{0}
 \end{cases}.
\end{aligned}
\label{Eq:BiFunc}
\end{equation}
}
\end{linenomath}
The bifurcating function has five parameters: $a_{0,1,2,3,4}$.

The values of \Rpf are measured in three regions defined by the $\eta$ of the \PQb quark candidate jet: $\abs{\eta}<0.5$, $0.5\leq \abs{\eta} <1.4$, $1.4\leq \abs{\eta} <2.4$, and are multiplied by the event yield in the regions SR$^\prime$ and VR$^\prime$ to obtain the multijet background in the SR and VR, respectively. 
This is expressed in Eq.~\eqref{Eq:qcd_yeild}, where $f^{1}_{\text{p/f}}$ and $f^{2}_{\text{p/f}}$ represent the fitted functions for $R^{1}_{\text{p/f}}$ and $R^{2}_{\text{p/f}}$, respectively.
\begin{linenomath}
\begin{equation}
\begin{split}
& \text{Multijet background in SR} = f^{1}_{\text{p/f}} (\mtb) \ \text{SR}^\prime; \\
& \text{Multijet background in VR} = f^{2}_{\text{p/f}} (\mtb) \ \text{VR}^\prime. 
\end{split}
\label{Eq:qcd_yeild}
\end{equation}
\end{linenomath}
The value of $f^{1}_{\text{p/f}}$ varies from 10 to 15\% in 2016 and 2 to 6\% from 2017 onwards, and the value of $f^{2}_{\text{p/f}}$ ranges from 3 to 12\% in 2016 and 1 to 2\% from 2017 onwards. 
Simulated \ttbar and single top quark backgrounds are subtracted from the data to calculate the yields in all regions.

In both data and simulation, it is observed that for AK4 jets that pass the threshold on the \PQb tagging discriminator, the associated AK8 jets have a different shape for the \msd distribution compared to the case where AK4 jets fail the \PQb tagging condition. 
This can affect the multijet \mtb spectrum extrapolated from the regions SR$^\prime$ and VR$^\prime$, leading to discrepancies with the SR and VR, respectively.
To take this effect into account, multiplicative corrections are applied to Eq.~\eqref{Eq:qcd_yeild}. For the estimate of the multijet background in the SR, the correction is derived using the ratio of \msd spectra of the AK8 jets associated with the \PQb quark candidate in CR1 and CR1$^\prime$, and  for the background in the VR it is obtained from CR2 and CR2$^\prime$.

The multijet background estimation procedure is first performed on a simulated QCD multijet sample, and the estimated \mtb distribution is obtained in the regions SR, VR, CR1, and CR2.
The difference between the extrapolated and predicted \mtb spectra in simulation is taken as a systematic uncertainty.

A closure test is performed in the VR by comparing the estimated multijet \mtb spectrum to that observed in data, after subtracting simulated \ttbar and single top quark backgrounds. 
The same test is performed in CR1 and CR2.
The predicted and observed distributions agree within 1--2\% in all cases.

To check the consistency of the simulated \ttbar background with data, a control region is selected
that satisfies all of the criteria in the SR, but requires that the AK8 jet associated with the \PQb quark candidate jet has \msd in the $[105, 210]\GeV$ window and passes the threshold on the \DeepAKX tagger score. 
This region is orthogonal to all of the regions specified in Table~\ref{tab:Region} 
and is enriched in \ttbar events where both the top quarks decay hadronically, 
which constitute approximately 80\% of the events in this region.
The ratio between the \mtb spectra in data and simulation is fitted with a first-order polynomial to derive a correction that is applied to the simulated \ttbar background.
The statistical uncertainties in the linear-fit parameters are used to derive the systematic uncertainty in this data-based correction applied to the simulated \ttbar background.

\section{Systematic uncertainties}

We consider several sources of systematic uncertainty that cover experimental effects, uncertainties due to the extraction of the multijet background, and 
uncertainties in the predicted \ttbar and single top quark backgrounds.
These sources and their sizes in the SR are as follows:

\begin{itemize}

\item \textit{Fit to }\Rpf: 
The impact of the uncertainty in \Rpf on the estimated multijet background is computed using the covariance matrix of the fit parameters, and ranges from 2 to 8\%.

\item \textit{Choice of function describing }\Rpf: 
The difference between the multijet background estimated using the default second-order polynomial and that obtained with the bifurcating function is less than 1\%. The value of this difference is taken to be the associated uncertainty.

\item \textit{Closure}:
The uncertainty is taken to be the difference between the estimated multijet background obtained by using \Rpf in simulation and the predicted yield in simulation. It ranges from 4 to 8\%.

\item \textit{\PQb quark candidate \msd correction}: 
The uncertainty is taken to be half of the difference between the estimated multijet background with and without applying the kinematic correction based on the \msd of the AK8 jet associated with the \PQb quark candidate jet. It is up to 6\% in magnitude.

\item \textit{JES}: 
The uncertainty applies to both AK4 and AK8 jets, and affects all of the backgrounds and the $\Wprime$ boson signal. It is taken to be fully correlated between AK4 and AK8 jets, and has a value rising to 5\% at high \mtb.

\item \textit{JER}: 
The uncertainty is taken into account both for simulation-based backgrounds and signal. It is taken to be fully correlated between AK4 and AK8 jets, and has a value ranging up to 8\%.

\item \textit{\PQb tagging scale factor}: 
The uncertainty in the correction applied in simulation to match the shape of the \DeepJet discriminator in data and simulation has a value of up to 30\%.

\item \textit{\PQt tagging scale factor}: 
The uncertainty in the correction applied in the simulation to match the efficiency of the \DeepAKX discriminator in data and simulation has a value of 4\%.

\item \textit{Trigger timing correction}: 
During the 2016 and 2017 data taking, a gradual shift in the timing of the inputs of the ECAL hardware level trigger in the region of $\abs{\eta}>2.0$ caused a specific trigger inefficiency. 
To take this effect into account, correction factors are computed from data and applied to the simulated samples corresponding to the 2016 and 2017 data taking periods. 
The uncertainty in this correction is less than 5\% over the entire \mtb range.

\item \textit{Pileup}: 
The value of the $\Pp\Pp$ total inelastic cross section that is used in the simulation of pileup events is varied upwards and downwards from its assumed value of 69.2\unit{mb} by its uncertainty of 4.6\%~\cite{Sirunyan:2018nqx}. The difference in yields is taken as the pileup uncertainty, and is less than 5\%.

\item \textit{Scale}: 
The impact of missing higher-order terms in perturbative QCD is evaluated by changing the renormalization ($\mu_{\mathrm{R}}$) and factorization ($\mu_{\mathrm{F}}$) scales in simulation. 
A six-point scheme is used, where yields are obtained in simulation by scaling the pairs ($\mu_{\mathrm{R}}$, $\mu_{\mathrm{F}}$) by the following combinations: (1, 0.5), (0.5, 1), (1, 2), (2, 1), (0.5, 0.5), and (2, 2). 
The envelope of the variations as compared to the nominal choice of ($\mu_{\mathrm{R}}, \ \mu_{\mathrm{F}}) = (1, \ 1)$ is taken as the scale uncertainty.
This reaches up to 20\% for \ttbar and single top quark backgrounds.

\item \textit{PDF}: 
The parameters that characterize the PDFs are determined from different experiments by fitting the theory prediction to the experimental data.
Thus the PDF parameters are affected by the uncertainties from the experimental measurements,
modeling, and parameterization assumptions.
In this analysis, simulated \ttbar and single top quark backgrounds are obtained for each of one hundred replicas of the NNPDF3.1 NNLO PDF set, and the
PDF uncertainty is determined from the standard deviation of the yields in each bin of the \mtb distribution for the \ttbar and single top backgrounds.
For the $\Wprime$ boson signal sample, the PDF uncertainty is determined using forty-five eigenvectors of the Hessian matrix~\cite{Pumplin:2001ct}.
The PDF uncertainty size is within 5\% for \ttbar and single top quark backgrounds and increases up to 20\% for a $\Wprime$  boson signal of high mass.

\item \textit{\ttbar normalization and slope}: 
The uncertainties in the linear fit parameters used to describe the data-based correction for the \ttbar background range from 5\% at low--{\mtb} to 25\% at high--{\mtb}.

\item \textit{Integrated luminosity}: 
The individual integrated luminosities of the 2016, 2017, and 2018 data-taking periods have uncertainties in the range 2.3--2.5\%~\cite{CMS-PAS-LUM-17-001,CMS-PAS-LUM-17-004,CMS-PAS-LUM-18-002}.
The total Run~2 (2016--2018) integrated luminosity has an uncertainty of 1.8\%, where the improvement in precision reflects the (uncorrelated) time evolution of various systematic effects.

\item \textit{\ttbar cross section}: 
This results in an uncertainty of $3.9\%$~\cite{Sirunyan:2017uhy} in the normalization of the \ttbar background.

\item \textit{Single top quark cross section}: 
There is an uncertainty of $12.8\%$ in the single top quark production cross section, averaged between the production in the $t$ channel~\cite{Sirunyan:2018rlu} and production in association with a \PW boson~\cite{Sirunyan:2018lcp}. This results in a corresponding uncertainty in the normalization of the single top quark background.
 
\end{itemize}

Since the multijet background is dominant, the uncertainties in \Rpf and in JES are the dominant sources of systematic uncertainty.

Systematic uncertainties corresponding to the \ttbar cross section, single top quark cross section, scale, and PDF are taken to be fully correlated, 
and the uncertainty in the integrated luminosity is taken to be partially correlated between different years of data taking. 
All other systematic uncertainties are taken to be uncorrelated between the three data-taking eras.

\section{Results}

The signal and expected background \mtb distributions are compared with data, and a binned maximum-likelihood fit based on Poisson statistics is applied to measure the $\Wprime$ boson yield.
Each source of systematic uncertainty is treated as a nuisance parameter~\cite{Conway:2011in}. The nuisance parameters corresponding to the systematic uncertainties that affect only the normalization of the backgrounds and the signal are assumed to follow log-normal distributions, and those that affect the normalization as well as the shape are described by Gaussian distributions.
The expected number, $N_\text{expected}$, of $\Wprime$ boson signal events in an \mtb bin is given by
\begin{linenomath}
\begin{equation}
N_\text{expected} =  \sigWp \mathcal{B} \mathcal{L} \epsilon \mathcal{A},
\label{Eq:N_exp}
\end{equation}
\end{linenomath}
where $\sigWp$ is the production cross section of the $\Wprime$ boson, 
$\mathcal{B}$
is the branching fraction of a $\Wprime$ boson decaying to a top and a bottom quark, followed by the hadronic decay of the \PW boson in the top quark decay chain, $\mathcal{L}$ denotes the integrated luminosity of the data used, and $\epsilon$ and $\mathcal{A}$ are the signal detection efficiency and the geometric and kinematic acceptance, respectively.

\begin{figure*}[hbtp]
\begin{center}
\includegraphics[width=0.321\textwidth]{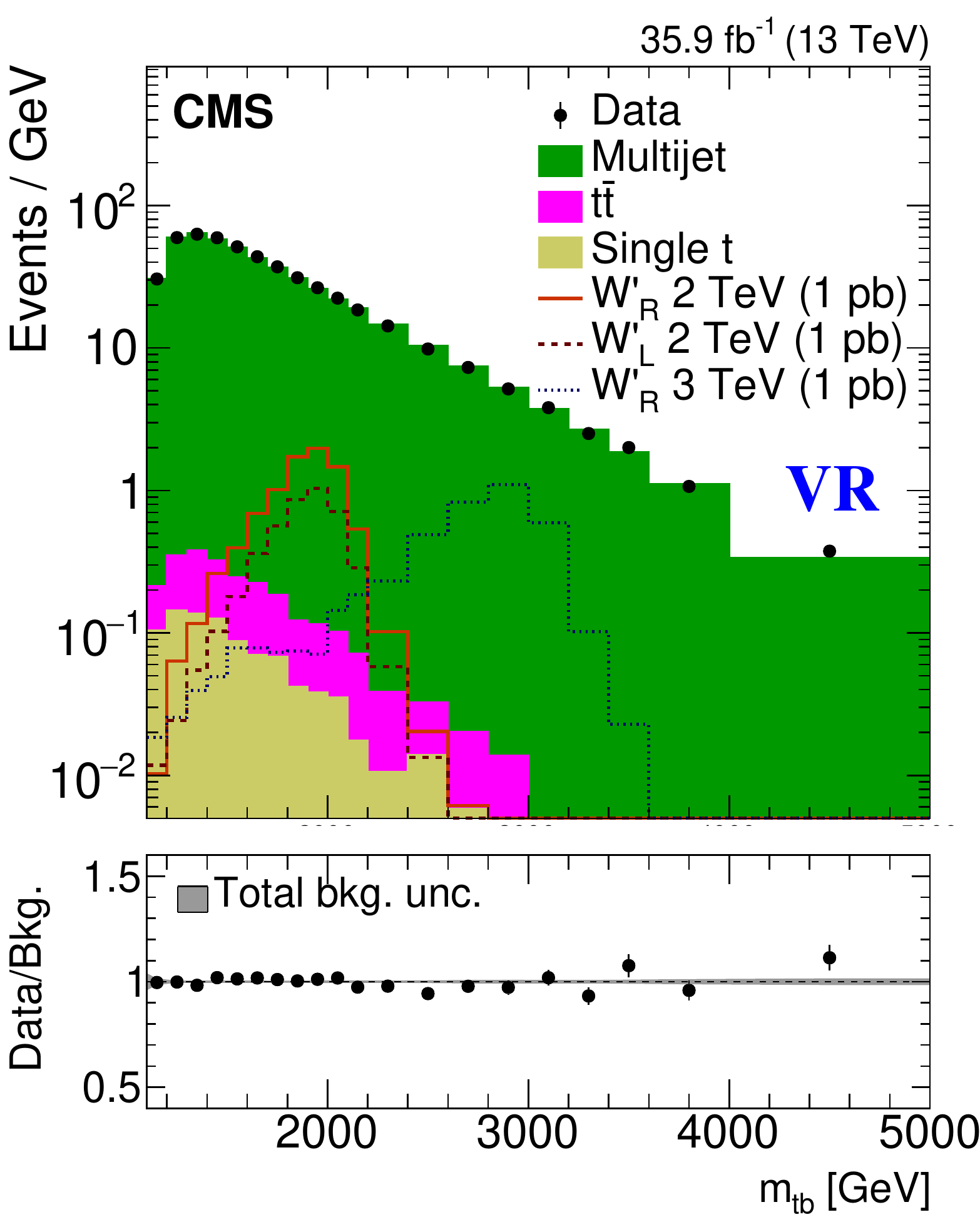}
\includegraphics[width=0.321\textwidth]{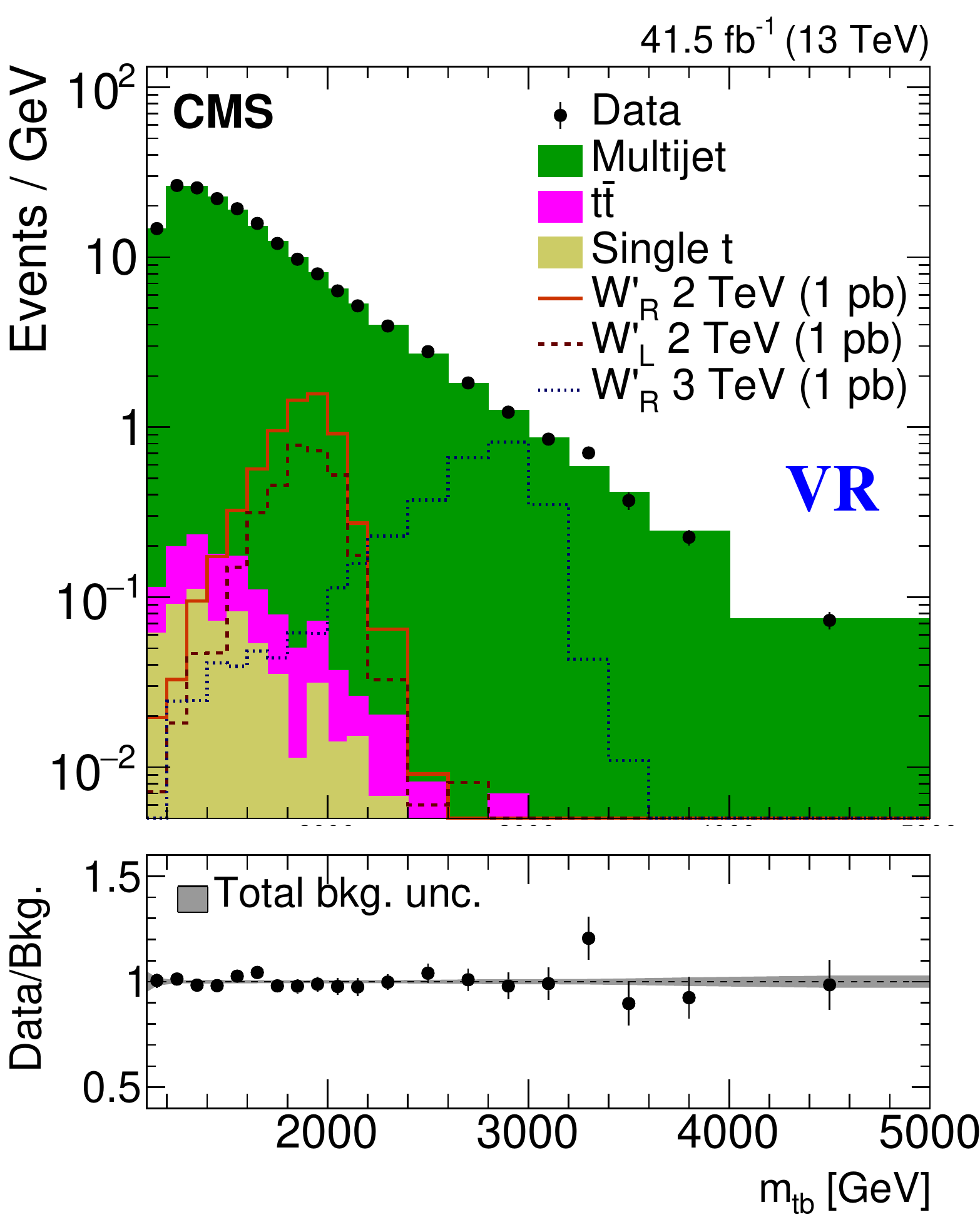}
\includegraphics[width=0.321\textwidth]{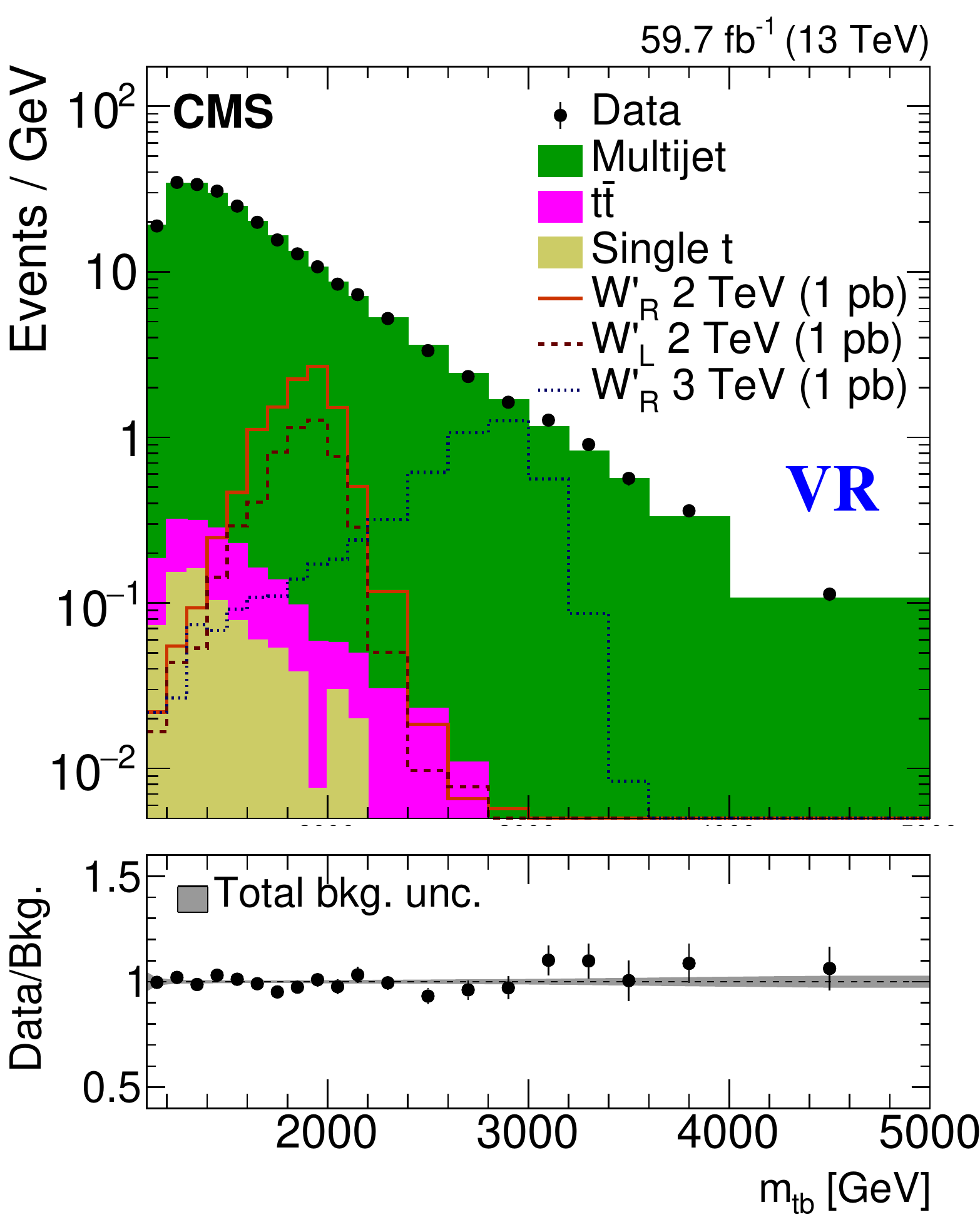}
\includegraphics[width=0.321\textwidth]{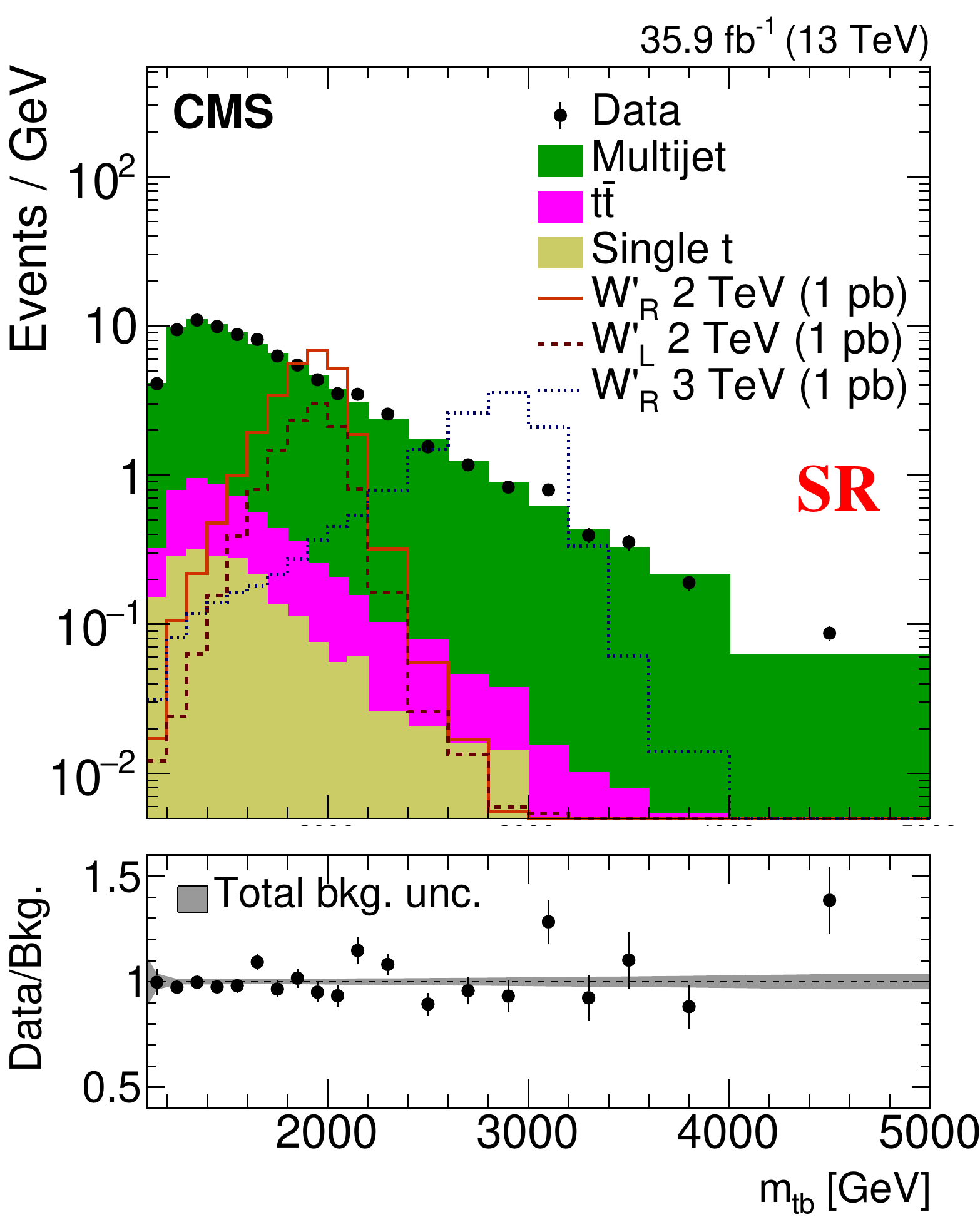}
\includegraphics[width=0.321\textwidth]{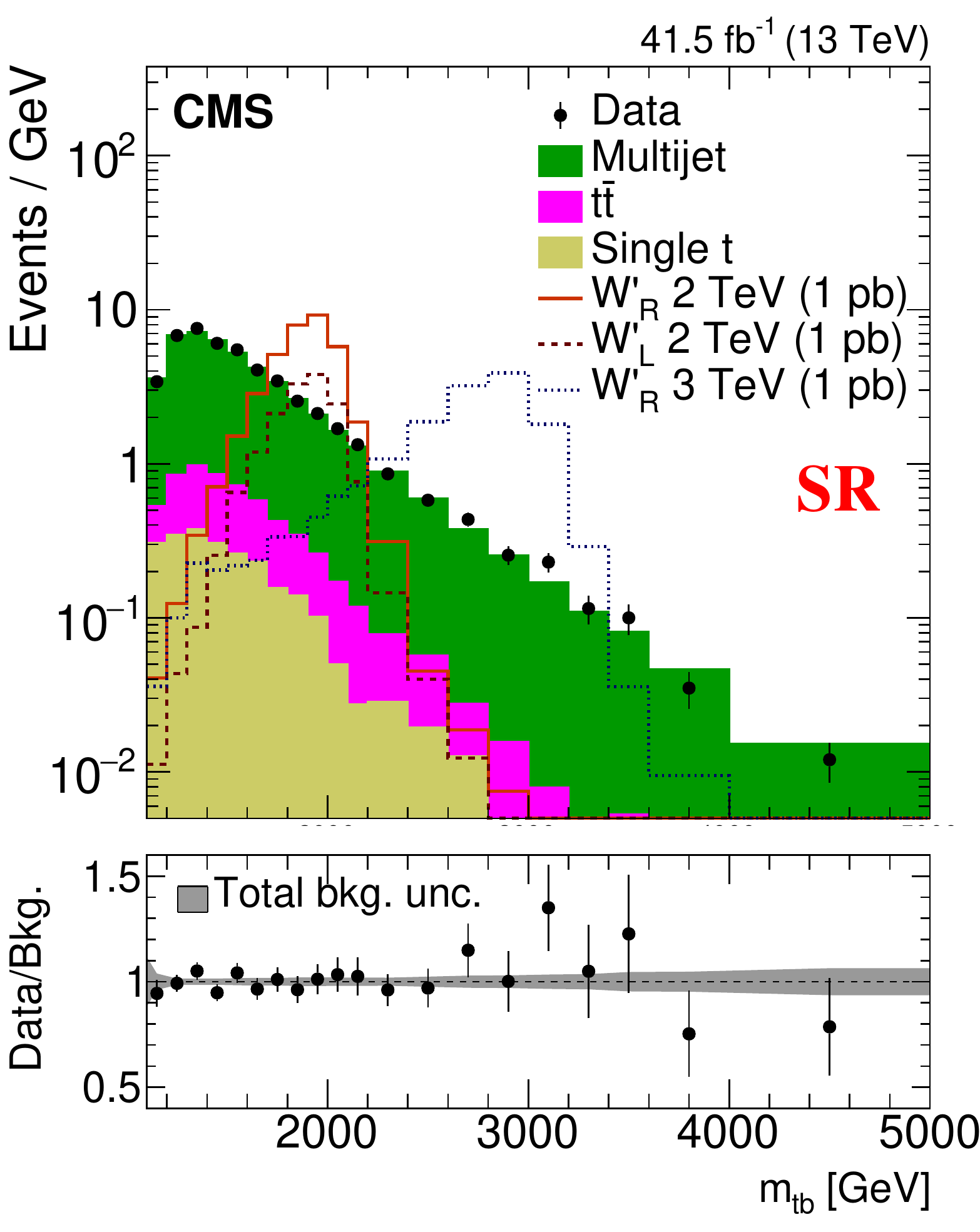}
\includegraphics[width=0.321\textwidth]{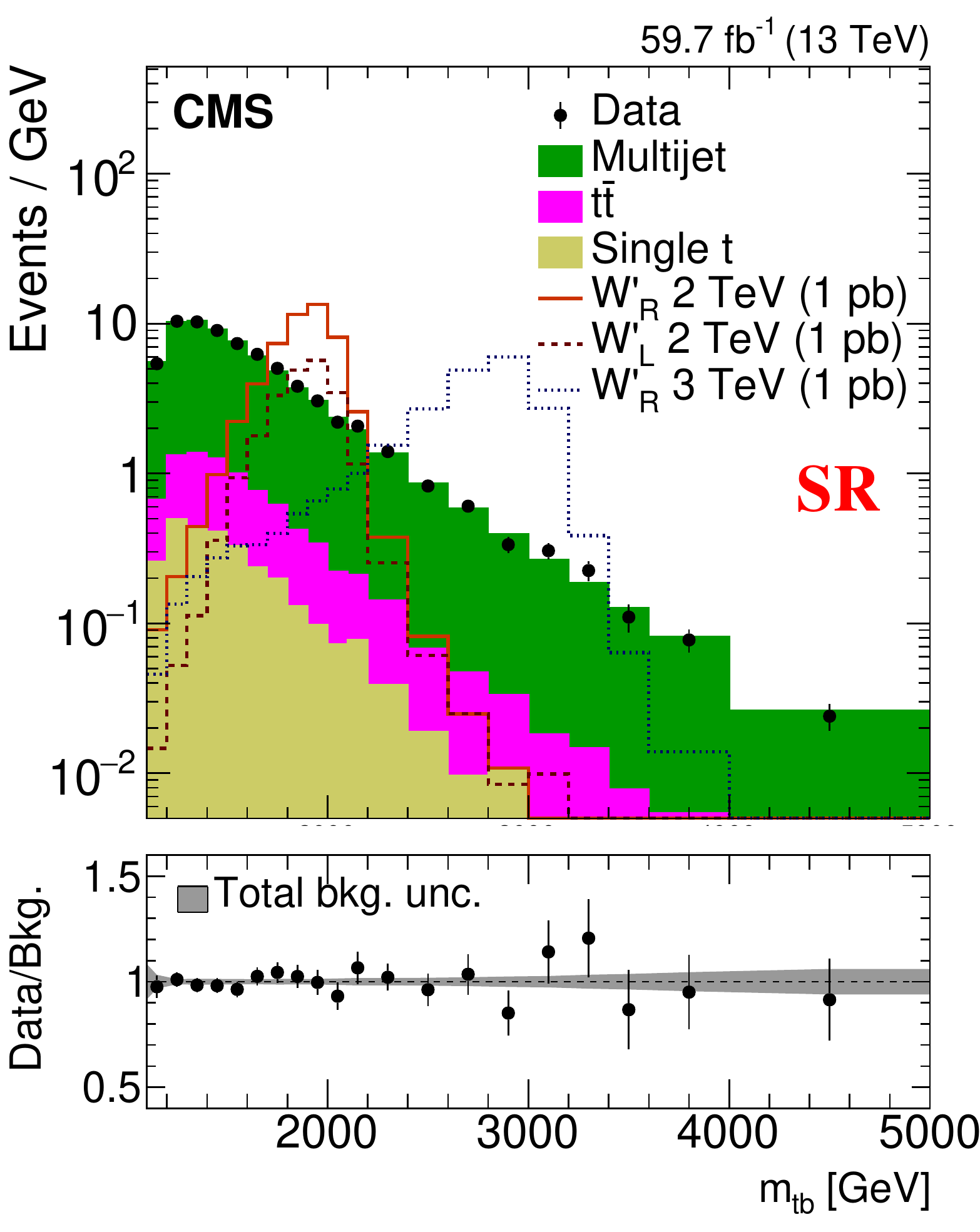}
\end{center}
\caption{
The reconstructed \mtb distributions in data (black points with error bars), and backgrounds in the VR (upper row) and SR (lower row) for the data-taking periods of 2016 (left), 2017 (middle), and 2018 (right). 
The yield in each bin is divided by the corresponding bin width.
Distributions expected from right-handed $\Wprime$ bosons of mass 2 and 3\TeV and a left-handed $\Wprime$ boson of mass 2\TeV are shown normalized to the integrated luminosity of the data using a product of cross section and branching fraction of 1\pb.
The lower panel in each plot shows the ratio of data to the background prediction. The shaded band indicates the total uncertainty in the estimated background, including both statistical and systematic components.
}
\label{fig:postfit}
\end{figure*}

The expected numbers of events from different background and signal hypotheses and the observed yields in data, after the binned maximum-likelihood fit is performed are shown in Fig.~\ref{fig:postfit} in the VR and SR for the three years of data taking. 
The agreement within the statistical and systematic uncertainties between the predicted SM background and the observed data in the VR validates the background estimation.
\begin{figure*}[hbtp]
\begin{center}
\includegraphics[width=0.8\textwidth]{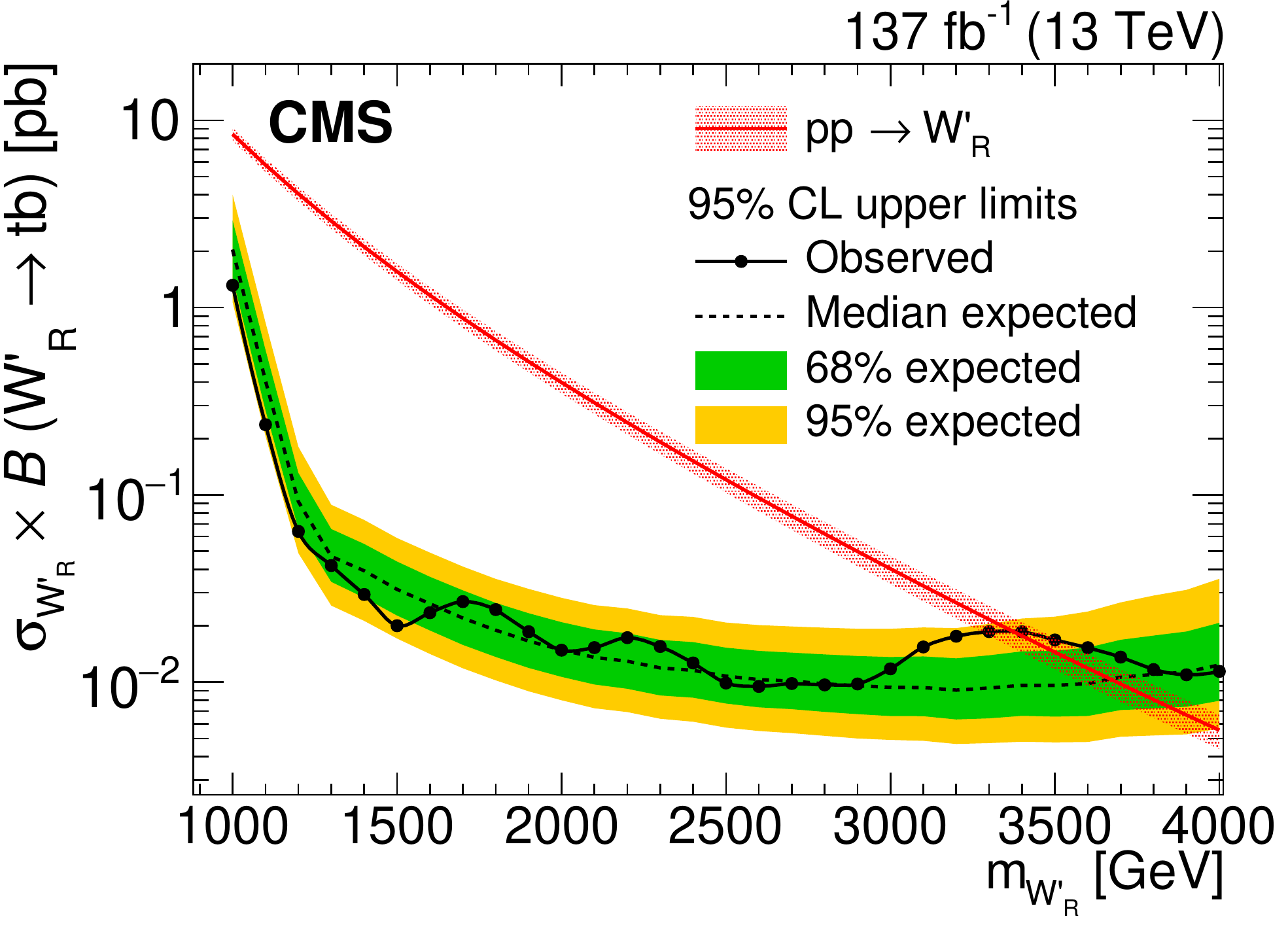}
\includegraphics[width=0.8\textwidth]{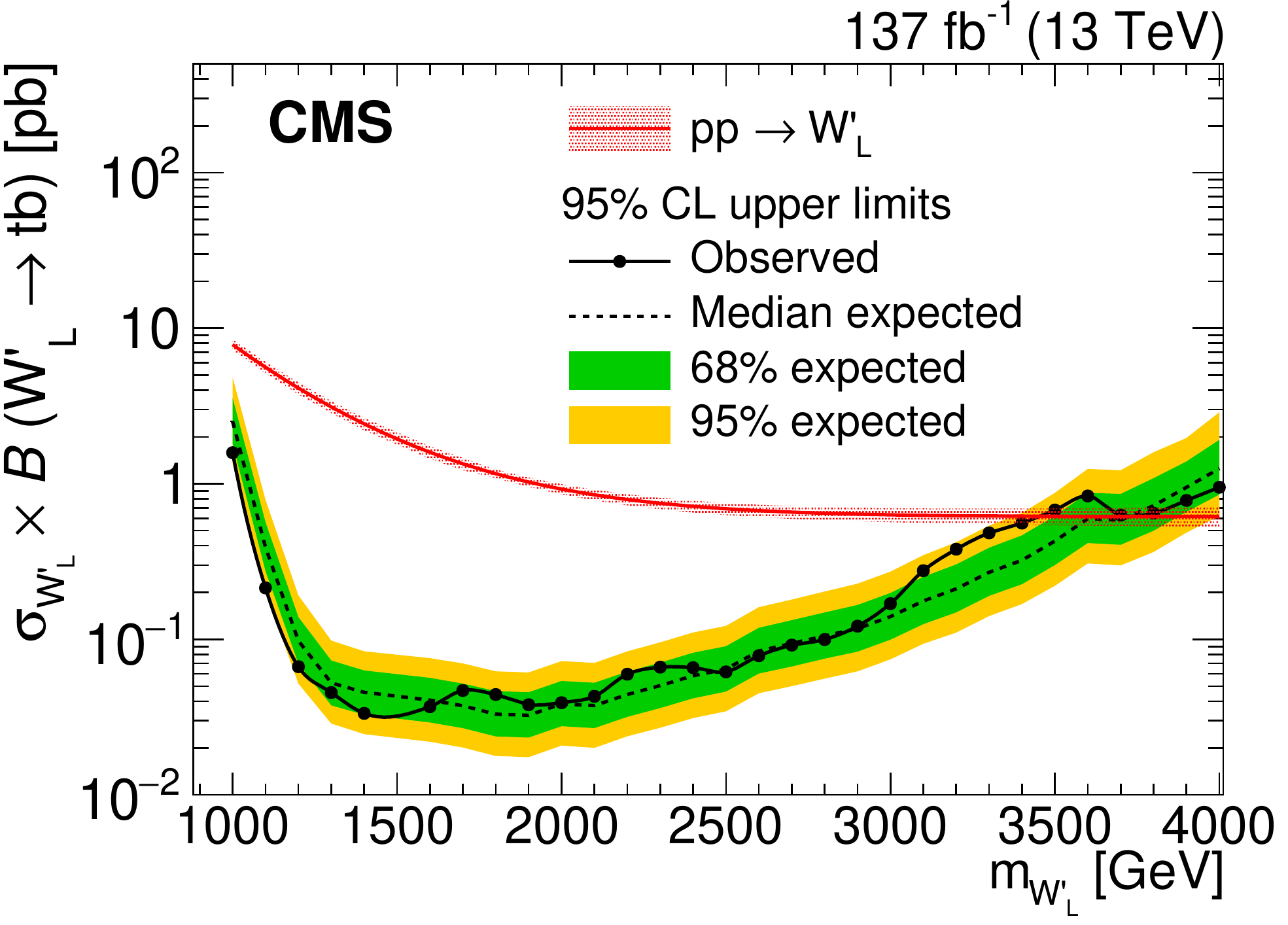}
\end{center}
\caption{
Upper limits at 95\% \CL on the production cross section and branching fraction of 
a W$^{\prime}_{\mathrm{R}}$ boson (upper row)
and a W$^{\prime}_{\mathrm{L}}$ boson with the SM interference (lower row)
decaying to a top and a bottom quark,
using combined 2016--2018 data and backgrounds.
The observed and median expected
limits are shown with the black solid and dashed lines, respectively.
The inner green and outer yellow bands represent the 68 and 95\% confidence level intervals, respectively, of the expected limit, computed using the background-only hypothesis.
The theoretical prediction and its uncertainty due to the choice of QCD scale and PDF set are indicated by the red curve and associated red shaded band, respectively.
}
\label{fig:limit}
\end{figure*}

No significant excess is observed over the SM background in the SR. 
Upper limits on  $\sigWp \mathcal{B}(\Wprime\to\PQt\PQb)$ 
at 95\% \CL
are obtained using the asymptotic \CLs method~\cite{Junk:1999kv,Read_2002}
with an asymptotic approximation~\cite{Cowan:2010js} of the profile likelihood.

Upper limits at 95\% \CL on the production cross sections times the branching fraction to a top and a bottom quark of right- and left-handed $\Wprime$ bosons including the effects of interference with the SM are calculated after merging the data and backgrounds of all three years and are shown in Fig.~\ref{fig:limit}.
There is a difference in the angular distributions of the top quark decay products depending on the chirality of the parent $\Wprime$ boson, which leads to a difference in \PQt tagging efficiency. 
The theoretical cross section for the production of left-handed $\Wprime$ bosons saturates at high mass because of the interference with single top quark production in the SM, 
which causes the signal shape to be asymmetric with a pronounced tail at low \mtb.
This results in a substantial difference between the upper limits on the production cross section of left- and right-handed $\Wprime$  bosons at high \mtb.

The current analysis excludes both right- and left-handed $\Wprime$ bosons of masses less than 3.4\TeV at 95\% \CL.  
The expected limits are 3.7 and 3.6\TeV for the right- and left-handed $\Wprime$ bosons, respectively.
The analysis improves on the expected sensitivity over the previous results~\cite{Sirunyan:2017vkm}, and extends the lower exclusion limit on $\sigWp \mathcal{B}(\Wprime\to\PQt\PQb)$ over the explored mass spectrum between 1.2 and 4.0\TeV. 
For right-handed $\Wprime$ bosons,
values of $\sigWp \mathcal{B}(\Wprime\to\PQt\PQb)$ in the range 60--10\;fb are excluded in the 1.2--4.0\TeV mass range.

\section{Summary}

A search has been performed for heavy $\Wprime$ bosons decaying to a top and a bottom quark in the hadronic final state using data corresponding to an integrated luminosity of 137\fbinv collected by the CMS experiment during the data taking period from 2016 to 2018.
The analysis utilizes top quark tagging and bottom quark tagging algorithms based on deep neural networks.
No excess above the estimated standard model background is observed.
Upper limits on the production cross section times branching fraction of a $\Wprime$ boson decaying to a top and a bottom quark are obtained at 95\% confidence level for $\Wprime$ boson masses in the range 1--4\TeV.
Left- and right-handed $\Wprime$ bosons with masses below 3.4\TeV are excluded at 95\% confidence level.
The limits provided on $\Wprime$ bosons decaying to a top and a bottom quark in the all-hadronic decay mode are the most stringent to date.

\begin{acknowledgments}

   We congratulate our colleagues in the CERN accelerator departments for the excellent performance of the LHC and thank the technical and administrative staffs at CERN and at other CMS institutes for their contributions to the success of the CMS effort. In addition, we gratefully acknowledge the computing centers and personnel of the Worldwide LHC Computing Grid and other centers for delivering so effectively the computing infrastructure essential to our analyses. Finally, we acknowledge the enduring support for the construction and operation of the LHC, the CMS detector, and the supporting computing infrastructure provided by the following funding agencies: BMBWF and FWF (Austria); FNRS and FWO (Belgium); CNPq, CAPES, FAPERJ, FAPERGS, and FAPESP (Brazil); MES (Bulgaria); CERN; CAS, MoST, and NSFC (China); MINCIENCIAS (Colombia); MSES and CSF (Croatia); RIF (Cyprus); SENESCYT (Ecuador); MoER, ERC PUT and ERDF (Estonia); Academy of Finland, MEC, and HIP (Finland); CEA and CNRS/IN2P3 (France); BMBF, DFG, and HGF (Germany); GSRT (Greece); NKFIA (Hungary); DAE and DST (India); IPM (Iran); SFI (Ireland); INFN (Italy); MSIP and NRF (Republic of Korea); MES (Latvia); LAS (Lithuania); MOE and UM (Malaysia); BUAP, CINVESTAV, CONACYT, LNS, SEP, and UASLP-FAI (Mexico); MOS (Montenegro); MBIE (New Zealand); PAEC (Pakistan); MSHE and NSC (Poland); FCT (Portugal); JINR (Dubna); MON, RosAtom, RAS, RFBR, and NRC KI (Russia); MESTD (Serbia); SEIDI, CPAN, PCTI, and FEDER (Spain); MOSTR (Sri Lanka); Swiss Funding Agencies (Switzerland); MST (Taipei); ThEPCenter, IPST, STAR, and NSTDA (Thailand); TUBITAK and TAEK (Turkey); NASU (Ukraine); STFC (United Kingdom); DOE and NSF (USA).
   
   \hyphenation{Rachada-pisek} Individuals have received support from the Marie-Curie program and the European Research Council and Horizon 2020 Grant, contract Nos.\ 675440, 724704, 752730, 765710 and 824093 (European Union); the Leventis Foundation; the Alfred P.\ Sloan Foundation; the Alexander von Humboldt Foundation; the Belgian Federal Science Policy Office; the Fonds pour la Formation \`a la Recherche dans l'Industrie et dans l'Agriculture (FRIA-Belgium); the Agentschap voor Innovatie door Wetenschap en Technologie (IWT-Belgium); the F.R.S.-FNRS and FWO (Belgium) under the ``Excellence of Science -- EOS" -- be.h project n.\ 30820817; the Beijing Municipal Science \& Technology Commission, No. Z191100007219010; the Ministry of Education, Youth and Sports (MEYS) of the Czech Republic; the Deutsche Forschungsgemeinschaft (DFG), under Germany's Excellence Strategy -- EXC 2121 ``Quantum Universe" -- 390833306, and under project number 400140256 - GRK2497; the Lend\"ulet (``Momentum") Program and the J\'anos Bolyai Research Scholarship of the Hungarian Academy of Sciences, the New National Excellence Program \'UNKP, the NKFIA research grants 123842, 123959, 124845, 124850, 125105, 128713, 128786, and 129058 (Hungary); the Council of Science and Industrial Research, India; the Ministry of Science and Higher Education and the National Science Center, contracts Opus 2014/15/B/ST2/03998 and 2015/19/B/ST2/02861 (Poland); the National Priorities Research Program by Qatar National Research Fund; the Ministry of Science and Higher Education, project no. 0723-2020-0041 (Russia); the Programa Estatal de Fomento de la Investigaci{\'o}n Cient{\'i}fica y T{\'e}cnica de Excelencia Mar\'{\i}a de Maeztu, grant MDM-2015-0509 and the Programa Severo Ochoa del Principado de Asturias; the Thalis and Aristeia programs cofinanced by EU-ESF and the Greek NSRF; the Rachadapisek Sompot Fund for Postdoctoral Fellowship, Chulalongkorn University and the Chulalongkorn Academic into Its 2nd Century Project Advancement Project (Thailand); the Kavli Foundation; the Nvidia Corporation; the SuperMicro Corporation; the Welch Foundation, contract C-1845; and the Weston Havens Foundation (USA).
\end{acknowledgments}

\bibliography{auto_generated}  

\cleardoublepage \appendix\section{The CMS Collaboration \label{app:collab}}\begin{sloppypar}\hyphenpenalty=5000\widowpenalty=500\clubpenalty=5000\vskip\cmsinstskip
\textbf{Yerevan Physics Institute, Yerevan, Armenia}\\*[0pt]
A.M.~Sirunyan$^{\textrm{\dag}}$, A.~Tumasyan
\vskip\cmsinstskip
\textbf{Institut f\"{u}r Hochenergiephysik, Wien, Austria}\\*[0pt]
W.~Adam, J.W.~Andrejkovic, T.~Bergauer, S.~Chatterjee, M.~Dragicevic, A.~Escalante~Del~Valle, R.~Fr\"{u}hwirth\cmsAuthorMark{1}, M.~Jeitler\cmsAuthorMark{1}, N.~Krammer, L.~Lechner, D.~Liko, I.~Mikulec, F.M.~Pitters, J.~Schieck\cmsAuthorMark{1}, R.~Sch\"{o}fbeck, M.~Spanring, S.~Templ, W.~Waltenberger, C.-E.~Wulz\cmsAuthorMark{1}
% ...
\vskip\cmsinstskip
\dag: Deceased\\
1:  Also at Vienna University of Technology, Vienna, Austria\\
% ...
\end{sloppypar}
\end{document}